\documentclass[10pt,conference]{IEEEtran}
\IEEEoverridecommandlockouts
\pdfoutput=1
\usepackage{graphics} % for pdf, bitmapped graphics files
\usepackage{epsfig} % for postscript graphics files
\usepackage{times} % assumes new font selection scheme installed
\usepackage{amsmath} % assumes amsmath package installed
\usepackage{algpseudocode}
\usepackage{algorithm}
\usepackage{flushend}

\allowdisplaybreaks[1]

\newtheorem{theorem}{Theorem}

\newcommand{\vect}[1]{\boldsymbol{#1}}
\newcommand{\figwidth}{0.85\columnwidth}

\makeatletter
\newcommand{\biggg}{\bBigg@{3}}
\newcommand{\vast}{\bBigg@{4}}
\newcommand{\Vast}{\bBigg@{5}}
\makeatother

\usepackage{color}

\begin{document}

\title{
Delay-Exponent of Bilayer Anytime Code}

\author{
\IEEEauthorblockN{Md. Noor-A-Rahim, Khoa D. Nguyen and Gottfried Lechner}
\IEEEauthorblockA{Institute for Telecommunications Research\\
University of South Australia\\
Adelaide, Australia\\
noomy004@mymail.unisa.edu.au, \{khoa.nguyen, gottfried.lechner\}@unisa.edu.au}
\thanks{This work was supported by the Australian Research Council Grant DE12010016.}
}

\maketitle

\maketitle
\thispagestyle{empty}
\pagestyle{empty}

%%%%%%%%%%%%%%%%%%%%%%%%%%%%%%%%%%%%%%%%%%%%%%%%%%%%%%%%%%%%%%%%%%%%%%%%%%%%%%%%
\begin{abstract}
In this paper, we study the design and the delay-exponent of anytime codes over a three terminal relay network.  We propose a bilayer anytime code based on anytime spatially coupled low-density parity-check (LDPC) codes and investigate the anytime characteristics through density evolution analysis. By using mathematical induction technique, we find analytical expressions of the delay-exponent for the proposed code. Through comparison, we show that the analytical delay-exponent has a close match with the delay-exponent obtained from numerical results.
\end{abstract}

\section{Introduction}
Anytime transmission was shown to be necessary and sufficient for tracking and controlling an unstable plant over a noisy channel \cite{1661825}. The necessary requirements of anytime transmission are causal encoding, decoding at any time instant and exponential decay of error probability with the decoding delay. Anytime transmission feasibility for the point-to-point case was first proved with sequential random codes in \cite{sahai2001anytime}. However, these codes were not suitable for practical implementation. More recently, practical anytime codes based on LDPC convolutional codes were proposed in \cite{Leefke} and \cite{our_anytime}.

In contrast to point-to-point anytime transmission, not much work has been done on anytime transmission over relay channels. Theoretical analysis of the delay-exponent for decode-and-forward (DF) relaying and compress-and-forward (CF) relaying were investigated in \cite{khoaDF} and \cite{6284045}, respectively. The first practical bilayer anytime codes for DF relaying were shown in our previous work \cite{AusCTW2014}. These bilayer codes offer simple encoding and provide good performance over relay channels. However, due to the multi edge type structure, the density evolution analysis for these bilayer codes is complicated. Moreover, no analysis for the delay-exponent was presented in \cite{AusCTW2014}.

The contributions of this work are as follows. We design a new bilayer anytime code for DF relaying which allows tractable analytical analysis. We show the anytime characteristics of the proposed code over binary erasure relay channels through density evolution analysis. We derive an analytical bound for the corresponding delay-exponent using mathematical induction technique and observe a close match with the numerical delay-exponent.

%The rest of the paper is organized as follows. In Section~\ref{sec:background}, we describe anytime communication, spatially coupled RA codes and the bilayer coding technique for DF relaying. In Section~\ref{sec:anytimeRA}, we propose the new anytime code design based on SC-RA codes. We  show asymptotic analysis, simulation results and compare them with existing results. We propose and analyse bilayer anytime codes for relay channels in Section~\ref{sec:bilayerRA}. Finally, we conclude in Section~\ref{sec:conclusion}.

\section{Preliminaries}\label{sec:background}

\subsection{Anytime Transmission}\label{sec:AnytimeTx}
 Consider a streaming source, where at time $t$, a $K$-bit message $\vect{m}_t \in \{0, 1\}^K$ is produced. At time instant $t$, the anytime channel encoder produces a $M$-bit channel input $\vect{u}_t \in \{0, 1\}^M$ as a function of the available source information $\vect{m}_1^t = [\vect{m}_1,\vect{m}_2,....,\vect{m}_t]$. We define the anytime channel encoder by the function $E_t(\vect{m}_1^t)$. The encoded message $\vect{u}_t$ is then transmitted over a noisy channel. The anytime channel decoder receives $\vect{v}_t$ (noisy version of $\vect{u}_t$) at every time step $t$. The decoder estimates the transmitted messages using the decoding function $\vect{\hat m}_1^t(t)=D_t(\vect{v}_1^t)$, i.e., the decoder produces an estimate $\vect{\hat m}_1^t(t)$ based on the current as well as all previously received messages. %A block diagram of anytime transmission is illustrated in Fig.~\ref{fig:anychannel}.
%\begin{figure}[htbp]
%\centering
%\includegraphics{anytime_channel.pdf}\\
%\caption{A block diagram illustrating anytime communication.}\label{fig:anychannel}
%\end{figure}

%\noindent
 We now consider the estimate $\vect{\hat m}_i(t)$ of message $\vect{m}_i$ at time $t\geq i$ ; the corresponding decoding delay for message $\vect{m}_i$ is $d=t-i$. The probability of error for the message $i$ at time $t$ will be $P_e (i,t) = \Pr\left(\vect{\hat{m}}_i(t) \neq \vect{m}_i\right)$. For a given channel, the encoder-decoder pair $(E_t,D_t)$ is called \emph{anytime code} if there exists a finite $\beta > 0$ such that \cite{simsek}:
\begin{equation}\label{eq:acc}
P_e (i,t) \leq \beta e^{-\alpha (t-i)},\hspace{0.2 cm} \forall \hspace{0.1 cm} (t-i)\geq 0,
\end{equation}
where $\alpha>0$ is known as \emph{delay-exponent}. The delay-exponent $\alpha$ specifies how fast the reliability of the system improves with the delay. The inequality in \eqref{eq:acc} is the major property of an anytime code. \eqref{eq:acc} implies that at any delay $d \geq 0$, the decoder is capable of estimating the message $\vect{m}_i$ and the probability of error for that message will decay exponentially to zero as the delay $d$ approaches infinity.

\subsection{Anytime Spatially Coupled LDPC Codes}
Anytime spatially coupled LDPC \cite{our_anytime} codes are constructed by coupling an infinite number of standard $(d_v,d_c)$-regular LDPC protographs, where $d_v$ and $d_c$ are the degrees of variable nodes and check nodes, respectively. We assume that each protograph/position contains $M$ variables and $\frac{d_v}{d_c}M$ check nodes. To ensure causal streaming and anytime decoding capability, each of the $d_v$ edges of a variable node at position $i$ is connected to a check node $j$ independently chosen from the range range $[i,i+1,...,i+\infty]$. The check node $j$ is chosen such that $j-i$ follows an exponential distribution and hence, a variable node at position $i$ has more connections with the check nodes close to position $i$. Particularly, the probability an edge originated from a variable node (VN) at position $i$ connects to a check node (CN) at position $j$ $(j\geq i)$ is
\begin{align}
P_r(k) = e^{-k\lambda}(1-e^{-\lambda}),
\end{align}
where $0<\lambda<1$ is known as exponential rate parameter and $k=j-i$ is the distance between the connected check node and variable node. In this paper, we refer to the above mentioned code as $(d_v,d_c,\lambda)$-anytime SC-LDPC code.

We consider $M$ to be the number of variable nodes at each position which represent a single message. We also consider $t$ as the number of messages received so far. For the density evolution of the above $(d_v,d_c,\lambda)$-anytime SCLDPC codes, we assume an infinite number of variable nodes $(M = \infty)$ at each position. Let $x^{(l)}(i,t)$ be the erasure probability of a message outgoing from a VN at position $i$ in the $l^{th}$ iteration at time $t$. For any $t\geq i$, we get the following density evolution equation for the $(d_v,d_c,\lambda)$-anytime SCLDPC code,
\noindent
\begin{align} \label{eq:OutErasure}
x^{(l)}(i,t) = &\epsilon\left(1-\sum\limits_{j=0}^{t-i}P_r(j)\bigg(1- \right. \nonumber \\
&\left.\sum\limits_{k=0}^{\infty}P_r(k)x^{(l-1)}(i+j-k,t)\bigg)^{d_c-1}\right)^{d_v-1},
\end{align}
where $ \epsilon$ is the channel erasure probability and
 \begin{align}
 x^{(0)}(i,t) =
 \begin{cases}
 0 & \text{if} \hspace{0.2 cm} i\leq 0 \\\nonumber
 \epsilon & \text{if} \hspace{0.2 cm} 0< i\leq t\\\nonumber
 1 & \text{if} \hspace{0.2 cm} i>t \nonumber
 \end{cases}
 \end{align}

\subsection{Relay Channel and Bilayer Codes}\label{sec:relaybilayer}
In this paper, we consider three terminal relay networks, which consist of a source, a relay and a destination. The communication links/channels of this system are: source-relay $sr$, relay-destination $rd$ and source-destination $sd$. We assume that all three links are binary erasure channels and are mutually orthogonal.
%Let the erasure probability of all links/channels are denoted as $\epsilon_{i}$ and $R_{i}$ $(i \in\{sr,rd,sd\})$, respectively.
Let $\epsilon_{sr}$, $\epsilon_{rd}$ and $\epsilon_{sd}$ denote the erasure probabilities of the three channels respectively.
As a relaying protocol, we consider the decode-and-forward strategy.

With a bilayer coding scheme \cite{bilayer}, the source first encodes the information bits using a code $\mathcal{C}_{sr}$ and then transmits the encoded bits. We assume that the relay can decode the message correctly. However, the destination is unable to recover the encoded bits received from the source due to the bad channel quality of the source-destination link. The relay generates extra parity bits based on the  received bits using a code $\mathcal{C}_r$. These extra parity bits are then encoded using another code $\mathcal{C}_{rd}$ and sent to destination. The destination recovers the source message with the help of the extra parity information received from the relay.

\section{Proposed Bilayer Anytime SC-LDPC Code}\label{sec:bilayer}
\subsection{Code design}\label{sec:code_design}
\begin{figure}[htbp]
  \centering
  \includegraphics[width=\figwidth]{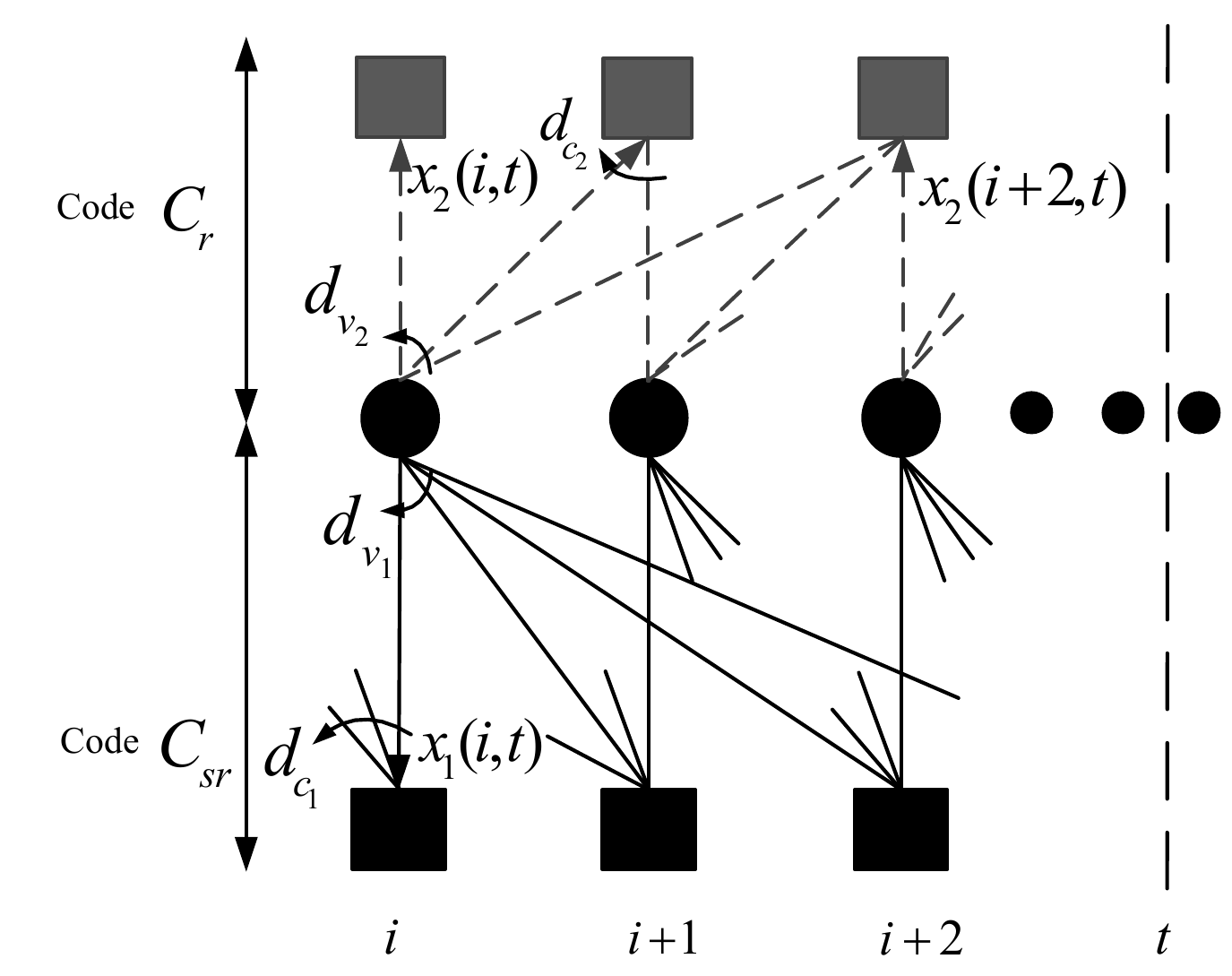}\\
  \caption{$(d_{v_1}, d_{c_1}, \lambda_1, d_{v_2}, d_{c_2}, \lambda_2)$-Bilayer anytime SC-LDPC code}\label{fig:bi_any}
\end{figure}

 We design the bilayer anytime codes for relay channels according to the bilayer coding scheme described in \cite{bilayer} and \cite{conv_bilayer}. The source causally encodes the messages by using a $(d_{v_1},d_{c_1},\lambda_1)$-anytime SC-LDPC code $\mathcal{C}_{sr}$ and broadcasts the encoded messages to the relay and destination. Due to the anytime characteristics of code $\mathcal{C}_{sr}$, the relay has to wait for a certain number of messages to receive for successful recovery of the source bits.  However, in this work, we ignore this initial delay required at the relay and hence we assume error-free transmission from the source to the relay. Based on the received messages, the relay causally generates extra parity bits using another $(d_{v_2},d_{c_2},\lambda_2)$-anytime SC-LDPC code $\mathcal{C}_r$.  Then the relay encodes these extra parity bits using a capacity approaching\footnote{ We refer to a code as capacity approaching code, when the rate of the code approaches the capacity of the corresponding channel. For BEC, the capacity of the channel $i$ is $C_{i} = 1 - \epsilon_{i}$, where $ \epsilon_{i}$ is the erasure probability of channel $i$.} block code for the relay destination link, which ensures secure transmission of relaying bits to the destination. The destination first decodes the extra parity bits from the relay and uses them as side information to decode the information received from the source. We refer to the combination of the code $\mathcal{C}_{sr}$ and the code $\mathcal{C}_{r}$ as $(d_{v_1}, d_{c_1}, \lambda_1, d_{v_2}, d_{c_2}, \lambda_2)$-bilayer anytime SC-LDPC code. Such a bilayer anytime code is depicted in Fig.~\ref{fig:bi_any}.
\subsection{Density Evolution}
Now we consider density evolution for the proposed bilayer anytime code.  Let $x_1^{(l)}(i,t)$ be the erasure probability of a message outgoing from a VN at position $i$ to the CNs of code $\mathcal{C}_{sr}$ in the $l^{th}$ iteration, at time $t$. Let $x_2^{(l)}(i,t)$ be the erasure probability of a message outgoing from a VN at position $i$ to the CNs of code $\mathcal{C}_r$ in the $l^{th}$ iteration, at time $t$. We also consider $P_{r_1}(k) = e^{-k\lambda_1}(1-e^{-\lambda_1})$ and $P_{r_2}(k) = e^{-k\lambda_2}(1-e^{-\lambda_2})$. At iteration $l$, we get the following update equations for $x_1^{(l)}(i,t)$ and $x_2^{(l)}(i,t)$:

\begin{align}\label{eq:out_x_1}
&x_1^{(l)}(i,t) = \epsilon_{sd}\left(1-\sum\limits_{j=0}^{t-i}P_{r_1}(j)\Bigg(1- \sum\limits_{k=0}^{\infty}P_{r_1}(k) \right. \nonumber\\
&\hspace{.1cm}. \; \left. x^{(l-1)}_1(i+j-k,t)\Bigg)^{d_{c_1}-1}\right)^{d_{v_1}-1}  \left(1-\sum\limits_{j=0}^{t-i}P_{r_2}(j) \right.\nonumber\\
&\hspace{.2cm}.\;\Bigg(1-  \left. \sum\limits_{k=0}^{\infty}P_{r_2}(k)x^{(l-1)}_2(i+j-k,t)\Bigg)^{d_{c_2}-1}\right)^{d_v{_2}},
\end{align}

\begin{align}\label{eq:out_x_2}
&x_2^{(l)}(i,t) = \epsilon_{sd}\left(1-\sum\limits_{j=0}^{t-i}P_{r_1}(j)\Bigg(1- \sum\limits_{k=0}^{\infty}P_{r_1}(k) \right. \nonumber\\
&\hspace{.1cm}.\; \left. x^{(l-1)}_1(i+j-k,t)\Bigg)^{d_{c_1}-1}\right)^{d_{v_1}} \left(1-\sum\limits_{j=0}^{t-i}P_{r_2}(j) \right. \nonumber\\
&\hspace{.2cm}.\; \Bigg(1-  \left. \sum\limits_{k=0}^{\infty}P_{r_2}(k)x^{(l-1)}_2(i+j-k,t)\Bigg)^{d_{c_2}-1}\right)^{d_v{_2}-1},
\end{align}

where
\begin{align}
 x_1^{(0)}(i,t) = x_2^{(0)}(i,t)=
 \begin{cases}
 0 & \text{if} \hspace{0.2 cm} i\leq 0 \\\nonumber
 \epsilon_{sd} & \text{if} \hspace{0.2 cm} 0< i\leq t\\\nonumber
 1 & \text{if} \hspace{0.2 cm} i>t \nonumber
 \end{cases}
 \end{align}
 The probability that VN $i$ has been erased can be computed by considering all the edges connected to VN $i$. Thus, the erasure probability $P_e (i,t)$ of message $i$ at decoding time instant $t$ is given by:
\begin{align}\label{eq:erasureProb}
&P_e(i,t) = \epsilon_{sd}\left(1-\sum\limits_{j=0}^{t-i}P_{r_1}(j)\Bigg(1- \sum\limits_{k=0}^{\infty}P_{r_1}(k) \right. \nonumber\\
&\hspace{.1cm}.\; \left. x^{(\infty)}_1(i+j-k,t)\Bigg)^{d_{c_1}-1}\right)^{d_{v_1}} \left(1-\sum\limits_{j=0}^{t-i}P_{r_2}(j) \right. \nonumber\\
&\hspace{.2cm}.\; \Bigg(1-  \left. \sum\limits_{k=0}^{\infty}P_{r_2}(k)x^{(\infty)}_2(i+j-k,t)\Bigg)^{d_{c_2}-1}\right)^{d_v{_2}},
\end{align}

\begin{figure}[htbp]
 \centering
 \includegraphics[width=\figwidth]{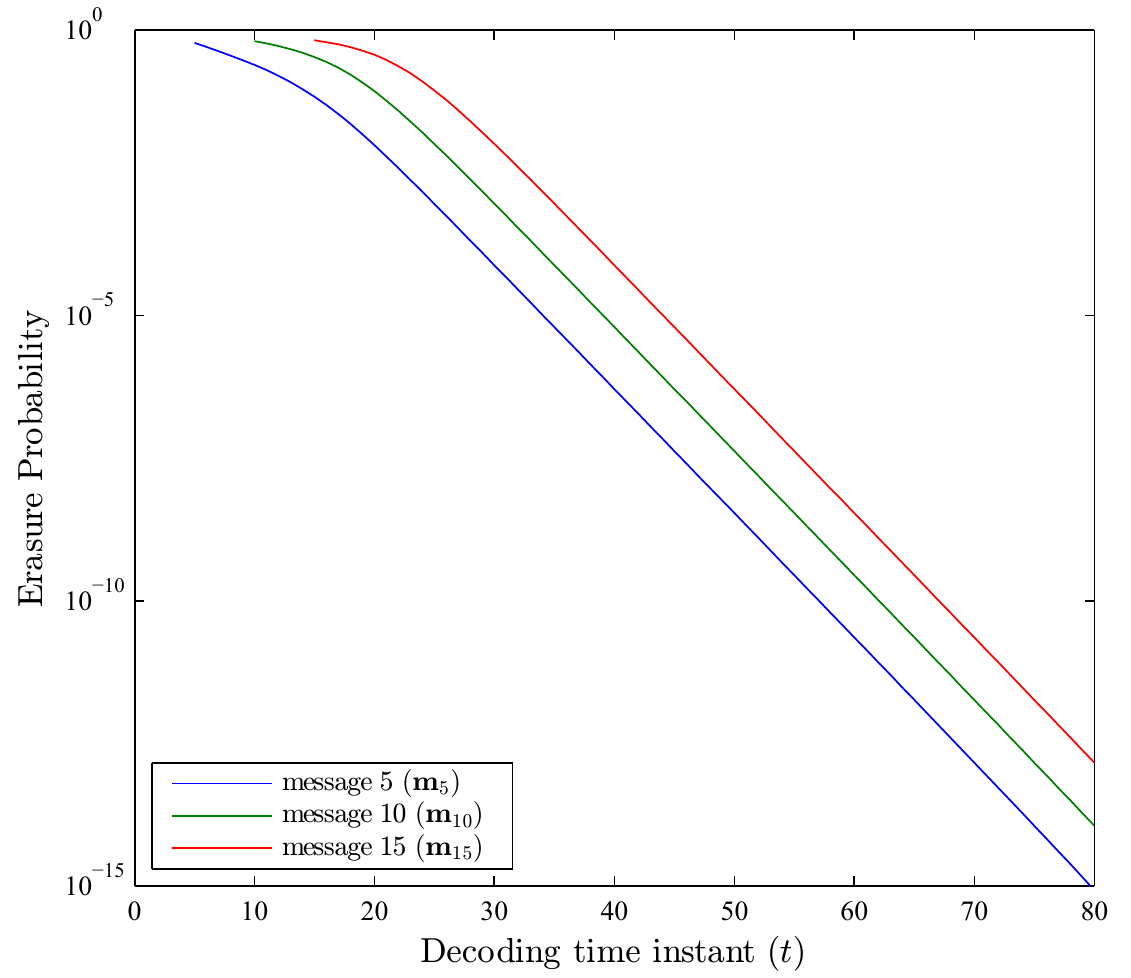}\\
 \caption{Asymptotic performance of the proposed bilayer anytime code over binary erasure channel ($\epsilon_{sd} = 0.7$).}\label{fig:asympshift}
\end{figure}

We present the results from the above density evolution analysis. We consider a $(d_{v_1}, d_{c_1}, \lambda_1, d_{v_2}, d_{c_2}, \lambda_2) = (3,6,0.1,2,8,0.1)$-bilayer anytime code.  Fig.~\ref{fig:asympshift} shows the erasure probabilities of three messages versus the decoding time instant. For each of the messages, we observe an exponential decaying of the erasure probability and the erasure performance of a given block is the shifted version of another. This shows that the error performance of a message is only dependent on the decoding delay and is independent on its position. From the erasure probability performance shown in Fig.~\ref{fig:asympshift}, numerically the delay-exponent $\alpha$ can be determined.

\subsection{Delay-Exponent Analysis}
The following theorem gives an analytical formula for the delay exponent.
\begin{theorem}\label{th:delayexponent}
Consider the $(d_{v_1}, d_{c_1}, \lambda_1, d_{v_2}, d_{c_2}, \lambda_2)$-bilayer anytime SC-LDPC code described in Subsection~\ref{sec:code_design}. For a source-destination channel erasure probability below a threshold $\bar{\epsilon}_{sd}$, asymptotically (i.e., for $M = \infty$) the delay-exponent of the proposed bilayer code is given by:
\begin{align}\label{eq:theoEx}
\alpha = \lambda_1 d_{v_1} + \lambda_2 d_{v_2},
\end{align}
where the threshold $\bar{\epsilon}_{sd}$ is defined by:

\begin{align}
&\bar{\epsilon}_{sd} =  \max\limits_{\beta>0} \; \min\limits_{d\geq 0} \; \beta \left(e^{-\lambda_1}+ K'\right)^{-d_{v_1}} \left(e^{-\lambda_2}+ K''\right)^{-d_{v_2}}\nonumber
\end{align}

%\begin{align}
%&\bar{\epsilon} =  \max\limits_{\beta>0} \; \min\limits_{d\geq 0} \; \beta \biggg(e^{-\lambda_1}+\beta (d_{c_1}-1) e^{\left(\lambda_1(2-d_{v_1})-\lambda_2 d_{v_2}\right)(d+1)} \nonumber\\
%&\; .\; (1-e^{-\lambda_1})\biggg(\frac{\left(1-e^{-\lambda_1}\right)\left(1-e^{\left(\lambda_1(d_{v_1}-2)+\lambda_2 d_{v_2}\right)(d+2)}\right)}{\left(1-e^{-\left(\lambda_1 d_{v_1}+\lambda_2 d_{v_2}\right)}\right)\left(1-e^{\lambda_1(d_{v_1}-2)+\lambda_2 d_{v_2}}\right)} + \nonumber\\
%&  \bigg(e^{\lambda_1 (d_{v_1}-1)+ \lambda_2 d_{v_2}}- \frac{1-e^{-\lambda_1}}{1-e^{-(\lambda_1 d_{v_1} + \lambda_2 d_{v_2})}}\bigg) \frac{1-e^{-2\lambda_1(d+2)}}{1-e^{-2\lambda_1}}\biggg)\biggg)^{-d_{v_1}}\nonumber\\
%& .\; \biggg(e^{-\lambda_2}+\beta (d_{c_2}-1) e^{\left(-\lambda_1 d_{v_1}+\lambda_2(2-d_{v_2})\right)(d+1)} (1-e^{-\lambda_2})  \nonumber\\
%& \biggg(\frac{(1-e^{-\lambda_2})\left(1-e^{\left(\lambda_1 d_{v_1}+\lambda_2(d_{v_2}-2)\right)(d+2)}\right)}{\left(1-e^{-\left(\lambda_1 d_{v_1}+\lambda_2 d_{v_2}\right)}\right)\left(1-e^{\lambda_1 d_{v_1}+\lambda_2(d_{v_2}-2)}\right)} + \nonumber\\
%&\left(e^{\lambda_1 d_{v_1}+\lambda_2 (d_{v_2}-1)}-\frac{1-e^{-\lambda_2}}{1-e^{-(\lambda_1 d_{v_1} +\lambda_2 d_{v_2})}}\right)\nonumber\\
%&\frac{1-e^{-2\lambda_2(d+2)}}{1-e^{-2\lambda_2}}\biggg)\biggg)^{-d_{v_2}}.\nonumber
%\end{align}

\end{theorem}

\begin{figure}[htbp]
 \centering
 \includegraphics[width=0.82\columnwidth]{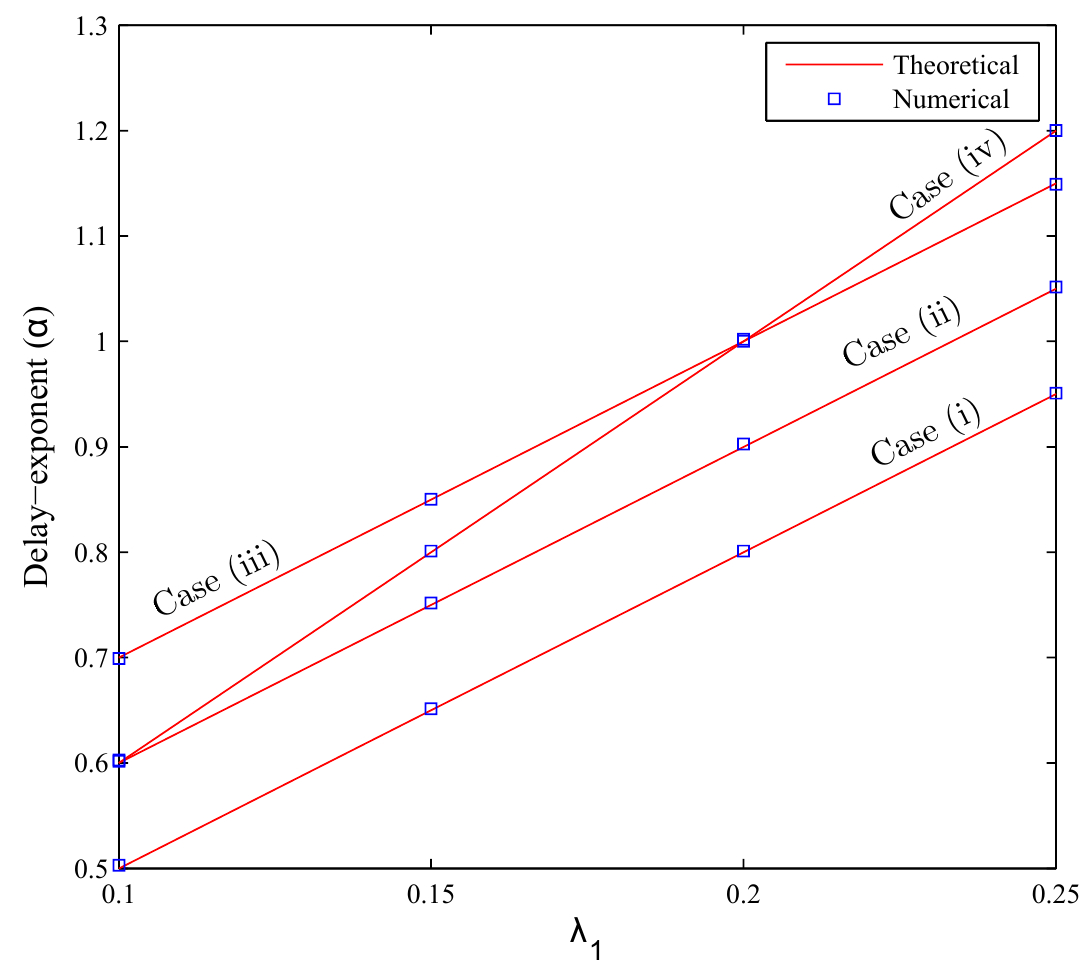}\\
 \caption{Comparison between theoretical and numerical delay-exponents for four cases: (i) $d_{v_1} = 3, d_{v_2} = 2, \lambda_2 = 0.1$ (ii) $d_{v_1} = 3, d_{v_2} = 3, \lambda_2 = 0.1$ (iii) $d_{v_1} = 3, d_{v_2} = 2, \lambda_2 = 0.2$ (iv) $d_{v_1} = 4, d_{v_2} = 2, \lambda_2 = 0.1$. For all cases, we consider $d_{c_1} = 6, d_{c_2} = 8$.}\label{fig:exponent}
\end{figure}

The proof of the theorem and the definitions of $K'$ and $K''$ are given in the appendix. In Fig.~\ref{fig:exponent}, we compare the theoretical exponent with the exponent obtained from the numerical results of density evolution. From Fig.~\ref{fig:exponent}, it is clear that the theoretically determined delay-exponent $\alpha = \lambda_1 d_{v_1} + \lambda_2 d_{v_2}$ closely matches the exponent of the decoded erasure probability obtained from density evolution analysis. We observe that the delay-exponent does not depend on the channel erasure probability $\epsilon_{sd}$ as long as $\epsilon_{sd}$ is below the threshold $\bar{\epsilon}_{sd}$. It is worth mentioning that there exists a large gap between the numerical and the theoretical thresholds due to the approximations taken in the proof of the theorem. For example, with a $(3,6,0.1,2,8,0.1)$-bilayer anytime SC-LDPC code, both the theoretical and numerical delay-exponent is $\alpha = 0.5$ while the theoretical and numerical thresholds are $0.115$ and $0.732$, respectively. From \eqref{eq:theoEx}, we observe that asymptotically the delay-exponent increases linearly with the increment of any parameter ($d_{v_1}$, $d_{v_2}$, $\lambda_1$ and $\lambda_2$). However, in practice (i.e., for finite length case), increasing $d_{v_1}$ or $d_{v_2}$  may introduce short cycles in the code. On the other hand, increasing $\lambda_1$ or $\lambda_2$ leads to fewer connections to distant protographs. Thus, increasing $d_{v_1}$, $d_{v_2}$, $\lambda_1$ or $\lambda_2$ may result in an error floor at large delays.

\section{Conclusion}\label{sec:conclusion}
In this paper, we proposed a practical bilayer anytime coding scheme for decode-and-forward relay channel. Through asymptotic analysis, we investigate the anytime characteristics of the proposed code, while ignoring the effect of delays at the source-relay link. For the first time, we have analytically found the delay exponent of a practical code over relay channels and observed a close prediction of the numerical delay-exponent. Our future work will be involved with the analysis of bilayer anytime codes while considering the effect of delays at the source-relay link.

\appendix[Proof of Theorem~\ref{th:delayexponent}]
At iteration $l=\infty$, \eqref{eq:out_x_1}, \eqref{eq:out_x_2} and \eqref{eq:erasureProb} can be combined as follows:%
\begin{align}\label{eq:de}
&x_e(i,t) = \epsilon_{sd}\left(1-\sum\limits_{j=0}^{t-i}P_{r_1}(j)\Bigg(1-  \right. \nonumber\\
&\hspace{.25cm}\; \left. \sum\limits_{k=0}^{\infty}P_{r_1}(k)x^{(l-1)}_1(i+j-k,t)\Bigg)^{d_{c_1}-1}\right)^{q_1}\nonumber\\
&\hspace{.5cm}.\; \left(1-\sum\limits_{j=0}^{t-i}P_{r_2}(j)\Bigg(1-  \right. \nonumber\\
&\hspace{.75cm}\; \left. \sum\limits_{k=0}^{\infty}P_{r_2}(k)x^{(l-1)}_2(i+j-k,t)\Bigg)^{d_{c_2}-1}\right)^{q_2},
\end{align}
where \eqref{eq:de} represents
\begin{itemize}
  \item  \eqref{eq:out_x_1} when $q_1=d_{v_1}-1$ and $q_2=d_{v_2}$.
  \item \eqref{eq:out_x_2} when $q_1=d_{v_1}$ and $q_2=d_{v_2}-1$.
  \item  \eqref{eq:erasureProb} when $q_1=d_{v_1}$ and $q_2=d_{v_2}$.
\end{itemize}
By using the following mathematical induction, we find the delay-exponent for the proposed bilayer  anytime code.
\begin{description}
  \item[\textbf{The basis}:] \hspace{.6cm}  For $0<i\leq d$,
  \begin{subequations}
\begin{align}
& \hspace{-1cm} x_1{(t-i,t)} \leq \beta e^{-\alpha_1 i}, \; \text{where}\; \alpha_1 = \lambda_1 (d_{v_1}-1) + \lambda_2 d_{v_2}\nonumber\\
& \hspace{-1cm} x_2{(t-i,t)} \leq \beta e^{-\alpha_2 i}, \; \text{where}\; \alpha_2 = \lambda_1 d_{v_1} + \lambda_2 (d_{v_2}-1)\nonumber\\
& \hspace{-1cm} P_e{(t-i,t)} \leq \beta e^{-\alpha i}, \; \text{where}\; \alpha = \lambda_1 d_{v_1} + \lambda_2 d_{v_2}\nonumber
\end{align}
\end{subequations}
  \item[\textbf{The inductive step}:] \hspace{1.9 cm}  We have to prove:
\begin{subequations}
\begin{align}
&x_1{(t-d-1,t)} \leq \beta e^{-\alpha_1(d+1)}\label{eq:proof_1}\\
&x_2{(t-d-1,t)} \leq \beta e^{-\alpha_2(d+1)}\label{eq:proof_2}\\
&P_e{(t-d-1,t)} \leq \beta e^{-\alpha(d+1)}\label{eq:proof_3}
\end{align}
\end{subequations}
\end{description}

In the following proof, we consider  $K_1 = e^{-k\lambda_1}(1-e^{-\lambda_1}) =  K_1 e^{-k\lambda_1}$ and $P_{r_2}(k) = e^{-k\lambda_2}(1-e^{-\lambda_2}) = K_2 e^{-k\lambda_2}$. Note that due to the page limitation, we skip few steps in the following proof. From \eqref{eq:de}, we get
\begin{align}\label{eq:Th_proof}
&x_e(t-d-1,t) = \epsilon_{sd}\biggg(1-\sum\limits_{j=0}^{d+1}P_{r_1}(j) \Bigg(1-\sum\limits_{k=0}^{\infty}P_{r_1}(k) \nonumber\\
&\hspace{.5cm}.\;x_1(t-d-1+j-k,t)\Bigg)^{d_{c_1}-1}\biggg)^{q_1} \biggg(1-\sum\limits_{j=0}^{d+1}P_{r_2}(j) \nonumber\\
&\hspace{.75cm}.\;\left.\left(1-\sum\limits_{k=0}^{\infty}P_{r_2}(k)x_2(t-d-1+j-k,t)\right)^{d_{c_2}-1}\right)^{q_2}\nonumber\\
&\overset{(a)}{\leq} \epsilon_{sd}\biggg(1-\sum\limits_{j=0}^{d+1}P_{r_1}(j)\Bigg(1-\beta\sum\limits_{k=0}^{j-1}P_{r_1}(k)  \nonumber\\
&\hspace{.5cm} .\; e^{-\alpha_1(d+1-j+k)} -\beta\sum\limits_{k=j}^{\infty}P_{r_1}(k)e^{-\alpha_1 d} \Bigg)^{d_{c_1}-1}\biggg)^{q_1}\nonumber\\
&\hspace{.75cm} .\; \biggg(1-\sum\limits_{j=0}^{d+1}P_{r_2}(j)\Bigg(1-\beta\sum\limits_{k=0}^{j-1}P_{r_2}(k)  \nonumber\\
&\hspace{1cm} .\; e^{-\alpha_2(d+1-j+k)} -\beta\sum\limits_{k=j}^{\infty}P_{r_2}(k)e^{-\alpha_2 d} \Bigg)^{d_{c_2}-1}\biggg)^{q_2}\nonumber\\
&= \epsilon_{sd}\biggg(1-\sum\limits_{j=0}^{d+1}P_{r_1}(j)\Bigg(1- \beta K_1    \nonumber\\
&\hspace{.5cm} .\; \sum\limits_{k=0}^{j-1}e^{-\lambda_1 k - \alpha_1(d+1-j+k)}-\beta e^{-\lambda_1 j-\alpha_1 d} \Bigg)^{d_{c_1}-1}\biggg)^{q_1} \nonumber\\
& \hspace{.75cm} .\; \biggg(1-\sum\limits_{j=0}^{d+1}P_{r_2}(j) \Bigg(1-\beta K_2  \nonumber\\
&\hspace{1cm} .\; \sum\limits_{k=0}^{j-1}e^{-\lambda_2 k - \alpha_2(d+1-j+k)}-\beta e^{-\lambda_2 j-\alpha_2 d} \Bigg)^{d_{c_2}-1}\biggg)^{q_2}\nonumber\\
&= \epsilon_{sd}\biggg(1-\sum\limits_{j=0}^{d+1}P_{r_1}(j)\Bigg(1- \beta K_1e^{-\alpha_1(d+1-j)}   \nonumber\\
&\hspace{.5cm} .\; \frac{1-e^{(-\lambda_1 - \alpha_1)j}}{1-e^{-\lambda_1 - \alpha_1}} -\beta e^{-\lambda_1 j-\alpha_1 d} \Bigg)^{d_{c_1}-1}\biggg)^{q_1}\nonumber\\
&\hspace{.75cm}.\; \biggg(1-\sum\limits_{j=0}^{d+1}P_{r_2}(j)\Bigg(1- \beta K_2e^{-\alpha_2(d+1-j)}   \nonumber\\
&\hspace{1cm}.\;  \frac{1-e^{(-\lambda_2 - \alpha_2)j}}{1-e^{-\lambda_2 - \alpha_2}} -\beta e^{-\lambda_2 j-\alpha_2 d} \Bigg)^{d_{c_2}-1}\biggg)^{q_2}\nonumber\\
&\overset{(b)}{\leq} \epsilon_{sd}\biggg(1-\sum\limits_{j=0}^{d+1}P_{r_1}(j)\Bigg(1-(d_{c_1}-1)  \Bigg(\beta K_1e^{-\alpha_1 (d+1-j)} \nonumber\\
&\hspace{.25cm}.\;  \frac{1-e^{(-\lambda_1 - \alpha_1)j}}{1-e^{-\lambda_1 - \alpha_1}} +\beta e^{-\lambda_1 j-\alpha_1 d}\Bigg)\Bigg)\biggg)^{q_1}\nonumber\\
&\hspace{.5cm} .\; \biggg(1-\sum\limits_{j=0}^{d+1}P_{r_2}(j)\Bigg(1-(d_{c_2}-1) \Bigg(\beta K_2e^{-\alpha_2 (d+1-j)} \nonumber\\
&\hspace{.75cm}.\; \frac{1-e^{(-\lambda_2 - \alpha_2)j}}{1-e^{-\lambda_2 - \alpha_2}} +\beta e^{-\lambda_2 j-\alpha_2 d}\Bigg)\Bigg)\biggg)^{q_2}\nonumber\\
&= \epsilon_{sd}\biggg(\sum\limits_{j=d+2}^{\infty}P_{r_1}(j)+\sum\limits_{j=0}^{d+1}P_{r_1}(j)(d_{c_1}-1)\Bigg(\beta K_1e^{-\alpha_1(d+1-j)} \nonumber\\
&\hspace{.25cm}.\; \frac{1-e^{(-\lambda_1 - \alpha_1)j}}{1-e^{-\lambda_1 - \alpha_1}} +\beta e^{-\lambda_1 j-\alpha_1 d} \Bigg)\biggg)^{q_1}\nonumber\\
& \hspace{.5 cm} .\; \biggg(\sum\limits_{j=d+2}^{\infty}P_{r_2}(j)+\sum\limits_{j=0}^{d+1}P_{r_2}(j)(d_{c_2}-1)\Bigg(\beta K_2e^{-\alpha_2(d+1-j)} \nonumber\\
&\hspace{.75cm}.\; \frac{1-e^{(-\lambda_2 - \alpha_2)j}}{1-e^{-\lambda_2 - \alpha_2}} +\beta e^{-\lambda_2 j-\alpha_2 d} \Bigg)\biggg)^{q_2} \nonumber\\
&= \epsilon_{sd}\biggg(e^{-\lambda_1(d+2)}+\sum\limits_{j=0}^{d+1}K_1e^{-\lambda_1 j}(d_{c_1}-1)\Bigg(\beta K_1e^{-\alpha_1(d+1-j)} \nonumber\\ &\hspace{.25cm}.\; \frac{1-e^{(-\lambda_1 - \alpha_1)j}}{1-e^{-\lambda_1 - \alpha_1}} +\beta e^{-\lambda_1 j-\alpha_1 d}\Bigg)\biggg)^{q_1}\nonumber\\
& \hspace{.5 cm} .\;\biggg(e^{-\lambda_2(d+2)}+\sum\limits_{j=0}^{d+1}K_2e^{-\lambda_2 j}(d_{c_2}-1)\Bigg(\beta K_2e^{-\alpha_2(d+1-j)} \nonumber\\ &\hspace{.75cm}.\; \frac{1-e^{(-\lambda_2 - \alpha_2)j}}{1-e^{-\lambda_2 - \alpha_2}} +\beta e^{-\lambda_2 j-\alpha_2 d}\Bigg)\biggg)^{q_2}\nonumber\\
& = \epsilon_{sd}\biggg(e^{-\lambda_1(d+2)}+ e^{-\alpha_1(d+1)}(d_{c_1}-1)\beta K_1 \nonumber\\
&\hspace{.25cm}.\; \biggg(\frac{K_1}{1-e^{-\lambda_1 - \alpha_1}}\frac{1-e^{(-\lambda_1+\alpha_1)(d+2)}}{1-e^{-\lambda_1+\alpha_1}} + \nonumber\\
&\hspace{.5cm} \left(e^{\alpha_1}-\frac{K_1}{1-e^{-\lambda_1 - \alpha_1}}\right)\frac{1-e^{-2\lambda_1(d+2)}}{1-e^{-2\lambda_1}}\biggg)\biggg)^{q_1}\nonumber\\
& \hspace{.75cm}.\; \biggg(e^{-\lambda_2(d+2)}+ e^{-\alpha_2(d+1)}(d_{c_2}-1)\beta K_2 \nonumber\\
& \hspace{1cm}.\; \biggg(\frac{K_2}{1-e^{-\lambda_2 - \alpha_2}}\frac{1-e^{(-\lambda_2+\alpha_2)(d+2)}}{1-e^{-\lambda_2+\alpha_2}} + \nonumber\\
&\hspace{1.25cm} \left(e^{\alpha_2}-\frac{K_2}{1-e^{-\lambda_2 - \alpha_2}}\right)\frac{1-e^{-2\lambda_2(d+2)}}{1-e^{-2\lambda_2}}\biggg)\biggg)^{q_2} \nonumber\\
& = \epsilon_{sd} e^{-(\lambda_1 q_1 + \lambda_2 q_2)(d+1)}\left(e^{-\lambda_1}+ K'\right)^{q_1}\left(e^{-\lambda_2}+ K''\right)^{q_2},
\end{align}
where
\begin{align}
\hspace{-.1cm} &K' = e^{(\lambda_1-\alpha_1)(d+1)}(d_{c_1}-1)\beta K_1 \biggg(\frac{K_1}{1-e^{-\lambda_1 - \alpha_1}} \nonumber\\
& \frac{1-e^{(-\lambda_1+\alpha_1)(d+2)}}{1-e^{-\lambda_1+\alpha_1}} + \left(e^{\alpha_1}-\frac{K_1}{1-e^{-\lambda_1 - \alpha_1}}\right) \frac{1-e^{-2\lambda_1(d+2)}}{1-e^{-2\lambda_1}}\biggg) \nonumber
\end{align}
and
\begin{align}
\hspace{-.1cm} &K'' = e^{(\lambda_2-\alpha_2)(d+1)}(d_{c_2}-1)\beta K_2 \biggg(\frac{K_2}{1-e^{-\lambda_2 - \alpha_2}} \nonumber\\
& \frac{1-e^{(-\lambda_2+\alpha_2)(d+2)}}{1-e^{-\lambda_2+\alpha_2}} + \left(e^{\alpha_2}-\frac{K_2}{1-e^{-\lambda_2 - \alpha_2}}\right) \frac{1-e^{-2\lambda_2(d+2)}}{1-e^{-2\lambda_2}}\biggg) \nonumber
\end{align}
In step (a), we apply induction assumption and the fact that $x_1(i+k,t)\geq x_1(i,t)$ and $x_2(i+k,t)\geq x_2(i,t)$ for all $k\geq 0$. In step (b), we apply $(1-z)^n\geq 1-nz$ for $|z|\leq 1$.
With $ \epsilon_{sd} \leq  \min\limits_{d} \; \beta \left(e^{-\lambda_1}+ K'\right)^{-q_1} \left(e^{-\lambda_2}+ K''\right)^{-q_2}$ and from \eqref{eq:Th_proof}, we can prove
\begin{itemize}
  \item \eqref{eq:proof_1}, if $q_1=d_{v_1}-1$ and $q_2=d_{v_2}$.
  \item \eqref{eq:proof_2}, if $q_1=d_{v_1}$ and $q_2=d_{v_2}-1$.
  \item \eqref{eq:proof_3}, if $q_1=d_{v_1}$ and $q_2=d_{v_2}$.
\end{itemize}

From the proof of \eqref{eq:proof_3}, we get the delay-exponent mentioned in \eqref{eq:theoEx}.

\bibliography{ITW2014}
\bibliographystyle{IEEEtran}

\end{document}